\documentclass[twocolumn,showpacs,preprintnumbers,showkeys,amsmath,amssymb]{revtex4}
\usepackage{graphicx}
\usepackage{amssymb}
\usepackage{amsmath}
\usepackage{bm}
\usepackage[flushleft]{caption2}

\begin{document}
\setcounter{page}{1}
\title{Invariants in electromagnetic and gravitational adjoint fields}
\author{Zihua Weng}
\email{xmuwzh@xmu.edu.cn.}
\affiliation{School of Physics and
Mechanical \& Electrical Engineering,
\\Xiamen University, Xiamen 361005, China}

\begin{abstract}
The paper discusses the impact of adjoint fields on the conservation
laws in the gravitational field and electromagnetic field, by means
of the characteristics of octonions. When the adjoint field can not
be neglected, it will cause the predictions to departure slightly
from the conservation laws, which include mass continuity equation,
charge continuity equation, and conservation of spin. The adjoint
field of electromagnetic field has an effect on conservation of
mass, and that of gravitational field on conservation of charge. The
inferences explain how the adjoint field influences some
conservation laws in the gravitational field and electromagnetic
field.
\end{abstract}



\maketitle

\section{INTRODUCTION}

The algebra of quaternions \cite{hamilton} can be used to describe
the scalar invariants and some conservation laws in the
gravitational field \cite{newton}. The algebra of octonions
\cite{cayley} can be used to demonstrate the scalar invariants in
the case for electromagnetic field \cite{maxwell} and gravitational
field, including conservation of mass \cite{lavoisier} and
conservation of energy \cite{lorentz}. The results are only dealt
with quaternion operator, but octonion operator. In the octonion
space, the operator should be extended from the quaternion operator
to the octonion operator.

Making use of the octonion operator, the gravitational field
demonstrated by the octonion operator will generate an adjoint
field. The source of adjoint field includes the adjoint mass and
adjoint linear momentum. In general, the adjoint mass and its
movement can not be observed by usual experiments. However, when the
adjoint mass is combined with the ordinary mass to become one sort
of particles, their movements will be accompanied by some mechanical
effects. Moreover, this kind of adjoint mass may influence the
distribution of electric charge.

With the invariant property of octonions, we find that the adjoint
mass and field strength have the influence on conservation laws in
the gravitational field, under the octonion coordinate
transformation.

\section{Octonion transformation}

In the quaternion space for the gravitational field, the basis
vector is $\mathbb{E}_g$ = ($1$, $\emph{\textbf{i}}_1$,
$\emph{\textbf{i}}_2$, $\emph{\textbf{i}}_3$), and radius vector is
$\mathbb{R}_g$ = ($r_0$, $r_1$, $r_2$, $r_3$), with velocity
$\mathbb{V}_g$ = ($v_0$, $v_1$, $v_2$, $v_3$). For the
electromagnetic field, the basis vector is $\mathbb{E}_e$ =
($\emph{\textbf{I}}_0$, $\emph{\textbf{I}}_1$,
$\emph{\textbf{I}}_2$, $\emph{\textbf{I}}_3$), the radius vector is
$\mathbb{R}_e$ = ($R_0$, $R_1$, $R_2$, $R_3$), and the velocity is
$\mathbb{V}_e$ = ($V_0$, $V_1$, $V_2$, $V_3$).

The $\mathbb{E}_e$ is independent of the $\mathbb{E}_g$, with
$\mathbb{E}_e$ = $\mathbb{E}_g$ $\circ$ $\emph{\textbf{I}}_0$. The
basis vectors $\mathbb{E}_g$ and $\mathbb{E}_e$ can be combined
together to become the basis vector $\mathbb{E}$ of the octonion
space.
\begin{eqnarray}
\mathbb{E} = \mathbb{E}_g + \mathbb{E}_e = (1, \emph{\textbf{i}}_1,
\emph{\textbf{i}}_2, \emph{\textbf{i}}_3, \emph{\textbf{I}}_0,
\emph{\textbf{I}}_1, \emph{\textbf{I}}_2, \emph{\textbf{I}}_3)
\end{eqnarray}

In the octonion space, the radius vector $\mathbb{R}$ is
\begin{eqnarray}
\mathbb{R} = \Sigma ( r_i \emph{\textbf{i}}_i ) + \Sigma ( R_i
\emph{\textbf{I}}_i)
\end{eqnarray}
and the velocity $\mathbb{V}$  is
\begin{eqnarray}
\mathbb{V} = \Sigma ( v_i \emph{\textbf{i}}_i ) + \Sigma ( V_i
\emph{\textbf{I}}_i)
\end{eqnarray}
where, $r_0 = v_0 t$; $v_0$ is the speed of light, $t$ denotes the
time; $R_0 = V_0 T$; $V_0$ is the speed of light-like, $T$ denotes a
time-like physical quantity; the $\circ$ denotes the octonion
multiplication. $i = 0, 1, 2, 3$, $j = 1, 2, 3$,
$\emph{\textbf{i}}_0 = 1$.

We may consider directly the quaternion space as the two-dimensional
complex space, and the octonion space as the four-dimensional
complex space.

In some special cases, the adjoint mass is combined with the
ordinary mass to become a sort of special particle. Therefore we can
measure its various characteristics, and have following relation
equation.
\begin{eqnarray}
R_i \emph{\textbf{I}}_i = r_i \emph{\textbf{i}}_i \circ
\emph{\textbf{I}}_0~;~ V_i \emph{\textbf{I}}_i = v_i
\emph{\textbf{i}}_i \circ \emph{\textbf{I}}_0~.
\end{eqnarray}

The octonion quantity $\mathbb{D} (d_0, d_1, d_2, d_3, D_0, D_1,
D_2, D_3 )$ is defined as follows.
\begin{eqnarray}
\mathbb{D} = d_0 + \Sigma (d_j \emph{\textbf{i}}_j) + \Sigma (D_i
\emph{\textbf{I}}_i)
\end{eqnarray}
where, $d_i$ and $D_i$ are all real.

When the coordinate system is transformed into the other, the
physical quantity $\mathbb{D}$ will be transformed into the new
octonion $\mathbb{D}' (d'_0 , d'_1 , d'_2 , d'_3 , D'_0 , D'_1 ,
D'_2 , D'_3 )$ .
\begin{equation}
\mathbb{D}' = \mathbb{K}^* \circ \mathbb{D} \circ \mathbb{K}
\end{equation}
where, $\mathbb{K}$ is the octonion, and $\mathbb{K}^* \circ
\mathbb{K} = 1$; $*$ denotes the conjugate of octonion; $\circ$ is
the octonion multiplication.

The octonion $\mathbb{D}$ satisfies the following equations.
\begin{eqnarray}
d_0 = d'_0
\end{eqnarray}

In the above equation, the scalar part $d_0$ is preserved during the
octonion coordinates are transforming. Some scalar invariants of
electromagnetic field will be obtained from this characteristics of
the octonion.

\begin{table}[t]
\caption{\label{tab:table1}The octonion multiplication table.}
\begin{ruledtabular}
\begin{tabular}{ccccccccc}
$ $ & $1$ & $\emph{\textbf{i}}_1$  & $\emph{\textbf{i}}_2$ &
$\emph{\textbf{i}}_3$  & $\emph{\textbf{I}}_0$  &
$\emph{\textbf{I}}_1$
& $\emph{\textbf{I}}_2$  & $\emph{\textbf{I}}_3$  \\
\hline $1$ & $1$ & $\emph{\textbf{i}}_1$  & $\emph{\textbf{i}}_2$ &
$\emph{\textbf{i}}_3$  & $\emph{\textbf{I}}_0$  &
$\emph{\textbf{I}}_1$
& $\emph{\textbf{I}}_2$  & $\emph{\textbf{I}}_3$  \\
$\emph{\textbf{i}}_1$ & $\emph{\textbf{i}}_1$ & $-1$ &
$\emph{\textbf{i}}_3$  & $-\emph{\textbf{i}}_2$ &
$\emph{\textbf{I}}_1$
& $-\emph{\textbf{I}}_0$ & $-\emph{\textbf{I}}_3$ & $\emph{\textbf{I}}_2$  \\
$\emph{\textbf{i}}_2$ & $\emph{\textbf{i}}_2$ &
$-\emph{\textbf{i}}_3$ & $-1$ & $\emph{\textbf{i}}_1$  &
$\emph{\textbf{I}}_2$  & $\emph{\textbf{I}}_3$
& $-\emph{\textbf{I}}_0$ & $-\emph{\textbf{I}}_1$ \\
$\emph{\textbf{i}}_3$ & $\emph{\textbf{i}}_3$ &
$\emph{\textbf{i}}_2$ & $-\emph{\textbf{i}}_1$ & $-1$ &
$\emph{\textbf{I}}_3$  & $-\emph{\textbf{I}}_2$
& $\emph{\textbf{I}}_1$  & $-\emph{\textbf{I}}_0$ \\
\hline $\emph{\textbf{I}}_0$ & $\emph{\textbf{I}}_0$ &
$-\emph{\textbf{I}}_1$ & $-\emph{\textbf{I}}_2$ &
$-\emph{\textbf{I}}_3$ & $-1$ & $\emph{\textbf{i}}_1$
& $\emph{\textbf{i}}_2$  & $\emph{\textbf{i}}_3$  \\
$\emph{\textbf{I}}_1$ & $\emph{\textbf{I}}_1$ &
$\emph{\textbf{I}}_0$ & $-\emph{\textbf{I}}_3$ &
$\emph{\textbf{I}}_2$  & $-\emph{\textbf{i}}_1$
& $-1$ & $-\emph{\textbf{i}}_3$ & $\emph{\textbf{i}}_2$  \\
$\emph{\textbf{I}}_2$ & $\emph{\textbf{I}}_2$ &
$\emph{\textbf{I}}_3$ & $\emph{\textbf{I}}_0$  &
$-\emph{\textbf{I}}_1$ & $-\emph{\textbf{i}}_2$
& $\emph{\textbf{i}}_3$  & $-1$ & $-\emph{\textbf{i}}_1$ \\
$\emph{\textbf{I}}_3$ & $\emph{\textbf{I}}_3$ &
$-\emph{\textbf{I}}_2$ & $\emph{\textbf{I}}_1$  &
$\emph{\textbf{I}}_0$  & $-\emph{\textbf{i}}_3$
& $-\emph{\textbf{i}}_2$ & $\emph{\textbf{i}}_1$  & $-1$ \\
\end{tabular}
\end{ruledtabular}
\end{table}

\section{Gravitational field}

The gravitational field and its adjoint field both can be
demonstrated by the quaternions, although they are quite different
from each other indeed.

\subsection{Invariants in gravitational field}

\subsubsection{Potential and strength}

The gravitational field potential is
\begin{eqnarray}
\mathbb{A}_g = \Sigma (a_i \emph{\textbf{i}}_i)~.
\end{eqnarray}

The strength $\mathbb{B}_g = \Sigma (b_{gi} \emph{\textbf{i}}_i) +
\Sigma (B_{gi} \emph{\textbf{I}}_i)$ consists of the gravitational
strength $\mathbb{B}_{gg}$ and adjoint field strength
$\mathbb{B}_{ge}$.
\begin{eqnarray}
\mathbb{B}_g = \lozenge \circ \mathbb{A}_g = \mathbb{B}_{gg} +
\mathbb{B}_{ge}
\end{eqnarray}
where, $\mathbb{B}_{gg} = \Sigma (b_{gi} \emph{\textbf{i}}_i)$,
$\mathbb{B}_{ge} = \Sigma (B_{gi} \emph{\textbf{I}}_i)$; $b_{gi}$
and $B_{gi}$ are all real; $ \lozenge = \Sigma \emph{\textbf{i}}_i (
\partial/\partial r_i) + \Sigma \emph{\textbf{I}}_i (
\partial/\partial R_i)$; $\partial_i =
\partial/\partial r_i$.

In the above equation, we choose the following gauge conditions to
simplify succeeding calculation.
\begin{eqnarray}
&& \partial a_0 / \partial r_0 - \Sigma (\partial a_j /
\partial r_j) = 0 \nonumber\\
&& \partial a_0 / \partial R_0 + \Sigma (\partial a_j /
\partial R_j) = 0 \nonumber
\end{eqnarray}

The gravitational strength $\mathbb{B}_{gg}$ in Eq.(9) includes two
parts, $\textbf{g}_g = ( g_{g01} , g_{g02} , g_{g03} ) $ and
$\textbf{b}_g = ( g_{g23} , g_{g31} , g_{g12} )$,
\begin{eqnarray}
\textbf{g}_g/v_0 = && \emph{\textbf{i}}_1 ( \partial_0 a_1 +
\partial_1 a_0 ) + \emph{\textbf{i}}_2 ( \partial_0 a_2 + \partial_2
a_0 )
\nonumber\\
&& + \emph{\textbf{i}}_3 ( \partial_0 a_3 + \partial_3 a_0 )
\\
\textbf{b}_g = && \emph{\textbf{i}}_1 ( \partial_2 a_3 -
\partial_3 a_2 ) + \emph{\textbf{i}}_2 ( \partial_3 a_1 - \partial_1
a_3 )
\nonumber\\
&& + \emph{\textbf{i}}_3 ( \partial_1 a_2 - \partial_2 a_1 )
\end{eqnarray}
while the adjoint field strength $\mathbb{B}_{ge}$ involves two
parts, $\textbf{E}_g = ( B_{g01} , B_{g02} , B_{g03} ) $ and
$\textbf{B}_g = ( B_{g23} , B_{g31} , B_{g12} )$ .
\begin{eqnarray}
\textbf{E}_g/v_0 = && \emph{\textbf{I}}_1 ( \partial_1 a_0 -
\partial_0 a_1 ) + \emph{\textbf{I}}_2 ( \partial_2 a_0 - \partial_0 a_2 )
\nonumber\\
&& + \emph{\textbf{I}}_3 ( \partial_3 a_0 - \partial_0 a_3 )
\\
\textbf{B}_g = && \emph{\textbf{I}}_1 ( \partial_3 a_2 -
\partial_2 a_3 ) + \emph{\textbf{I}}_2 ( \partial_1 a_3 - \partial_3
a_1 )
\nonumber\\
&& + \emph{\textbf{I}}_3 ( \partial_2 a_1 - \partial_1 a_2 )
\end{eqnarray}

The linear momentum density $\mathbb{S}_{gg} = m \mathbb{V}_g $ is
the source of gravitational field, and its adjoint linear momentum
density $\mathbb{S}_{ge} = \bar{m} \mathbb{V}_g \circ
\emph{\textbf{I}}_0$ is that of adjoint field. They combine together
to become the field source $\mathbb{S}_g$ .
\begin{eqnarray}
\mu \mathbb{S}_g && = - ( \mathbb{B}_g/v_0 + \lozenge)^* \circ
\mathbb{B}_g
\nonumber\\
&& = \mu_{gg} \mathbb{S}_{gg} + \mu_{ge} \mathbb{S}_{ge} -
\mathbb{B}_g^* \circ \mathbb{B}_g/v_0
\end{eqnarray}
where, $\bar{m}$ is the adjoint mass density; $\mu_{gg}$ and
$\mu_{ge}$ are the coefficients; $*$ denotes the conjugate of
octonion.

The $\mathbb{B}_g^* \circ \mathbb{B}_g/(2\mu_{gg})$ is the field
energy density.
\begin{eqnarray}
\mathbb{B}_g^* \circ \mathbb{B}_g/ \mu_{gg} = ( \mathbb{B}_{gg}^*
\circ \mathbb{B}_{gg} + \mathbb{B}_{ge}^* \circ \mathbb{B}_{ge} ) /
\mu_{gg}
\end{eqnarray}

The above means that the adjoint field energy makes a contribution
to the gravitational mass.

\subsubsection{Conservation of mass}

In the gravitational field and its adjoint field, the linear
momentum density $\mathbb{P} = \mu \mathbb{S}_g / \mu_{gg}$ is
written as
\begin{eqnarray}
\mathbb{P} = \widehat{m} v_0 + \Sigma (m v_j \emph{\textbf{i}}_j ) +
\Sigma ( M_g V_i \emph{\textbf{i}}_i \circ \emph{\textbf{I}}_0 )
\end{eqnarray}
where, $\widehat{m} = m - (\mathbb{B}_g^* \circ \mathbb{B}_g /
\mu_{gg} )/v_0^2 $ ; $M_g = \bar{m} \mu_{ge}/\mu_{gg}$ . $p_0 =
\widehat{m} v_0$, $p_j = m v_j $; $P_i = M_g V_i $.

The above means that the gravitational mass density $\widehat{m}$ is
changed with either the gravitational strength or the adjoint field
strength in the gravitational field and its adjoint field.

From Eq.(6), we have one linear momentum density, $\mathbb{P}'
(p'_0, p'_1, p'_2, p'_3, P'_0, P'_1, P'_2, P'_3)$, when the octonion
coordinate system is rotated. And we obtain the invariant equation
from Eqs.(7) and (16).
\begin{eqnarray}
\widehat{m} v_0 = \widehat{m}' v'_0
\end{eqnarray}

Under Eqs.(3), (7), and (17), we find the gravitational mass density
$\widehat{m}$ remains unchanged.
\begin{eqnarray}
\widehat{m} = \widehat{m}'
\end{eqnarray}

The above means that if we choose the definitions of velocity and
linear momentum, the inertial mass density and gravitational mass
density will keep unchanged respectively, under the coordinate
transformation in Eq.(6) in the gravitational field and its adjoint
field.

\subsubsection{Mass continuity equation}

In the gravitational field and its adjoint field, the applied force
density $\mathbb{F} = \Sigma (f_i \emph{\textbf{i}}_i ) + \Sigma
(F_i \emph{\textbf{I}}_i )$ is defined from the linear momentum
density $\mathbb{P}$ in Eq.(16).
\begin{eqnarray}
\mathbb{F} = v_0 (\mathbb{B}_g/v_0 + \lozenge )^* \circ \mathbb{P}
\end{eqnarray}
where, the scalar $f_0 = v_0 \Sigma (\partial p_i / \partial r_i) +
v_0 \Sigma ( \partial P_i / \partial R_i ) + \Sigma ( b_{gj} p_j +
B_{gj} P_j ) $.

When the coordinate system rotates, we have the new force density
$\mathbb{F}' (f'_0, f'_1, f'_2, f'_3, F'_0, F'_1, F'_2, F'_3)$.

By Eq.(7), we have
\begin{eqnarray}
f_0 = f'_0
\end{eqnarray}

When $f'_0 = 0$ in the above, we have the conservation of mass in
the case for coexistence of the gravitational field and its adjoint
field.
\begin{eqnarray}
\Sigma \left\{ \partial (p_i + P_i )/
\partial r_i \right\} + \Sigma ( b_{gj} p_j + B_{gj} P_j ) / v_0 = 0
\end{eqnarray}

If the $b_{gi} = B_{gi} = 0$, the above will be reduced to the
following equation.
\begin{eqnarray}
\partial (m + M_g )/ \partial t + \Sigma \left\{ \partial (p_j + P_j )/
\partial r_j \right\} = 0
\end{eqnarray}
further, if there is not adjoint field, we have
\begin{eqnarray}
\partial m / \partial t + \Sigma (\partial p_j / \partial r_j) = 0
\end{eqnarray}

The above states that the adjoint field strength, adjoint mass, and
gravitational strength have a tiny influence on the conservation of
mass in the gravitational field and its adjoint field, although the
$\Sigma ( b_{gj} p_j + B_{gj} P_j ) / v_0$ and $\Sigma ( \partial
P_i / \partial r_i )$ both are usually very tiny when the fields are
weak. In case of we choose the definitions of the applied force and
velocity in the gravitational field and adjoint field, the
conservation of mass will be the invariant under the octonion
transformation in Eq.(6).

\subsubsection{Conservation of spin}

The angular momentum density $\mathbb{L} = \Sigma (l_i
\emph{\textbf{i}}_i ) + \Sigma (L_i \emph{\textbf{I}}_i )$ is
defined from the radius vector $\mathbb{R}$, physical quantity
$\mathbb{X}$, and linear momentum density $\mathbb{P}$ in the
octonion space.
\begin{eqnarray}
\mathbb{L} = (\mathbb{R} + k_{rx} \mathbb{X}) \circ \mathbb{P}
\end{eqnarray}
where, $l_0$ is considered as the spin angular momentum density in
the gravitational field and adjoint field; $l_0 = (r_0 + k_{rx} x_0
) p_0 - \Sigma \left\{ (r_j + k_{rx} x_j ) p_j \right\} - \Sigma
\left\{ (R_i + k_{rx} X_i ) P_i \right\}$; $k_{rx}$ is the
coefficient.

When the octonion coordinate system rotates, we have the new angular
momentum density $\mathbb{L}' = \Sigma ( l'_i \emph{\textbf{i}}'_i +
L'_i \emph{\textbf{I}}'_i )$. Under the octonion coordinate
transformation, the spin density remains unchanged from Eq.(7).
\begin{eqnarray}
l_0 = l'_0
\end{eqnarray}

The above means the adjoint field, space, time, and strength have an
influence on orbital angular momentum and spin angular momentum. The
spin angular momentum density $l_0$ will change with time in the
gravitational field and adjoint field, although $l_0$ is one
invariant under the octonion transformation.

\begin{table}[b]
\caption{\label{tab:table1}The definitions and the mechanics
invariants of the gravitational field with its adjoint field in the
octonion space.}
\begin{ruledtabular}
\begin{tabular}{lll}
$definition$    &    $invariant $    &    $ meaning $                            \\
\hline
$\mathbb{R}$    &    $r_0 = r'_0 $   &    $Galilean~invariant$                   \\
$\mathbb{V}$    &    $v_0 = v'_0$    &    $invariable~speed~of~light$            \\
$\mathbb{A}$    &    $a_0 = a'_0$    &    $invariable~scalar~potential$          \\
$\mathbb{B}$    &    $b_0 = b'_0$    &    $invariable~gauge$                     \\
$\mathbb{P}$    &    $p_0 = p'_0$    &    $invariable~mass~density$              \\
$\mathbb{F}$    &    $f_0 = f'_0$    &    $conservation~of~mass$                 \\
$\mathbb{L}$    &    $l_0 = l'_0$    &    $invariable~spin~density$              \\
$\mathbb{W}$    &    $w_0 = w'_0$    &    $invariable~energy~density$            \\
$\mathbb{N}$    &    $n_0 = n'_0$    &    $conservation~of~energy$               \\
\end{tabular}
\end{ruledtabular}
\end{table}

\subsubsection{Conservation of energy}

The total energy density $\mathbb{W} = \Sigma (w_i
\emph{\textbf{i}}_i ) + \Sigma (W_i \emph{\textbf{I}}_i )$ is
defined from the angular momentum density $\mathbb{L}$.
\begin{eqnarray}
\mathbb{W} = v_0 ( \mathbb{B}_g/v_0 + \lozenge) \circ \mathbb{L}
\end{eqnarray}
where, the scalar part $w_0 = v_0 \partial l_0 / \partial r_0 - v_0
\Sigma (\partial l_j / \partial r_j) - v_0 \Sigma ( \partial L_i /
\partial R_i ) - \Sigma ( b_{gj} l_j + B_{gj} L_j ) $.

When the coordinate system rotates, we have the new energy density
$\mathbb{W}' = \Sigma ( w'_i \emph{\textbf{i}}'_i + W'_i
\emph{\textbf{I}}'_i )$. Under the octonion transformation, the
scalar part of total energy density is the energy density and
remains unchanged by Eq.(7).
\begin{eqnarray}
w_0 = w'_0
\end{eqnarray}

In some special cases, the right side is equal to zero. We obtain
the conservation of spin angular momentum.
\begin{eqnarray}
&& \partial l_0 / \partial r_0 - \Sigma (\partial l_j / \partial
r_j)
\nonumber\\
&& - \Sigma (\partial L_i / \partial r_i) - \Sigma ( b_{gj} l_j +
B_{gj} L_j ) / v_0 = 0
\end{eqnarray}

If the last term is neglected, the above is reduced to
\begin{eqnarray}
\partial l_0 / \partial r_0 - \Sigma (\partial l_j / \partial r_j)
- \Sigma ( \partial L_i /
\partial r_i ) = 0
\end{eqnarray}
further, if there is not adjoint field, we have
\begin{eqnarray}
\partial l_0 / \partial r_0 - \Sigma (\partial l_j / \partial
r_j) = 0
\end{eqnarray}

The above means the energy density $w_0$ is variable in the case for
coexistence of the gravitational field and adjoint field, because
the adjoint mass, velocity, and strength have the influence on the
angular momentum density. While the scalar $w_0$ is the invariant
under the octonion transformation from Eqs.(7) and (25).

\subsubsection{Conservation of power}

In the gravitational field with adjoint field, the external power
density $\mathbb{N}$  can be defined from the total energy density
$\mathbb{W}$ in Eq.(26).
\begin{eqnarray}
\mathbb{N} = v_0 ( \mathbb{B}_g/v_0 + \lozenge)^* \circ \mathbb{W}
\end{eqnarray}
where, the external power density $\mathbb{N}$ includes the power
density in the gravitational field and adjoint field.

The external power density can be rewritten as follows.
\begin{eqnarray}
\mathbb{N} = n_0 + \Sigma (n_j \emph{\textbf{i}}_j ) + \Sigma (N_i
\emph{\textbf{I}}_i )
\end{eqnarray}
where, the scalar $n_0 = v_0 \Sigma (\partial w_i / \partial r_i) +
v_0 \Sigma ( \partial W_i / \partial R_i ) + \Sigma ( b_{gj} w_j +
B_{gj} W_j ) $.

When the coordinate system rotates, we have the new external power
density $\mathbb{N}' = \Sigma ( n'_i \emph{\textbf{i}}'_i + N'_i
\emph{\textbf{I}}'_i )$. Under the octonion coordinate
transformation, the scalar part of external power density is the
power density and remains unchanged by Eq.(7).
\begin{eqnarray}
n_0 = n'_0
\end{eqnarray}

In a special case, the right side is equal to zero. And then, we
obtain the conservation of energy.
\begin{eqnarray}
\Sigma \left\{ \partial (w_i + W_i )/
\partial r_i \right\} + \Sigma ( b_{gj} w_j + B_{gj} W_j ) / v_0 = 0
\end{eqnarray}

If the last term is neglected, the above is reduced to
\begin{eqnarray}
\Sigma ( \partial w_i / \partial r_i ) + \Sigma ( \partial W_i /
\partial r_i ) = 0
\end{eqnarray}
further, if the last term is equal to zero, we have
\begin{eqnarray}
\Sigma ( \partial w_i / \partial r_i ) = 0
\end{eqnarray}

The above means that the power density $n_0$ will be variable in the
case for coexistence of the gravitational field and its adjoint
field, although the $n_0$ is the scalar invariant under the octonion
transformation. And the adjoint mass, strength, and torque density
etc. have a few influence on the energy continuity equation in the
gravitational field and the adjoint field.

\subsection{Invariants in gravitational adjoint field}

\subsubsection{Conservation of adjoint mass}

In the adjoint field, a new physical quantity $\mathbb{P}_g =
\mathbb{P} \circ \emph{\textbf{I}}_0^* $ can be defined from
Eq.(16).
\begin{eqnarray}
\mathbb{P}_g = M_g V_0 + \Sigma (M_g V_j \emph{\textbf{i}}_j ) -
\left\{ \widehat{m} v_0 \emph{\textbf{I}}_0 + \Sigma (m v_j
\emph{\textbf{I}}_j ) \right\}
\end{eqnarray}

By Eq.(6), we have the linear momentum density, $\mathbb{P}'_g =
\Sigma ( P'_i \emph{\textbf{i}}'_i - p'_i \emph{\textbf{I}}'_i )$,
when the octonion coordinate system is rotated. Under the octonion
coordinate transformation, the scalar part of $\mathbb{P}_g$ remains
unchanged.
\begin{eqnarray}
M_g V_0 = M'_g V'_0
\end{eqnarray}

With Eqs.(3), (7), and the above, we obtain the conservation of
adjoint mass as follows. And $M_g$ is the scalar invariant, which is
in direct proportion to the adjoint mass density $\bar{m}$ .
\begin{eqnarray}
M_g = M'_g
\end{eqnarray}

The above means that if we emphasize the definitions of velocity and
linear momentum, the adjoint mass density will remain the same,
under the coordinate transformation in the adjoint field and
gravitational field.

\subsubsection{Continuity equation of adjoint mass }

In the octonion space, a new physical quantity $\mathbb{F}_g =
\mathbb{F} \circ \emph{\textbf{I}}_0^*$ can be defined from Eq.(19).
\begin{eqnarray}
\mathbb{F}_g = F_0 + \Sigma (F_j \emph{\textbf{i}}_j ) - \Sigma (f_i
\emph{\textbf{I}}_i )
\end{eqnarray}
where, the scalar $F_0 = v_0 \Sigma ( \partial P_i / \partial r_i )
- v_0 \Sigma ( \partial p_i / \partial R_i ) + \Sigma ( b_{gj} P_j -
B_{gj} p_j ) $.

When the coordinate system rotates, we have the octonion applied
force density $\mathbb{F}'_g = \Sigma ( F'_i \emph{\textbf{i}}'_i -
f'_i \emph{\textbf{I}}'_i )$. Under the coordinate transformation,
the scalar part of $\mathbb{F}_g$ remains unchanged.
\begin{eqnarray}
F_0 = F'_0
\end{eqnarray}

When the right side is equal to zero in the above, we have the
continuity equation of adjoint mass in the case for coexistence of
the adjoint field and gravitational field.
\begin{eqnarray}
\Sigma \left\{ \partial ( P_i - p_i ) / \partial r_i \right\} +
\Sigma ( b_{gj} P_j - B_{gj} p_j ) / v_0 = 0
\end{eqnarray}

If the last term is neglected, the above is reduced to
\begin{eqnarray}
\Sigma ( \partial P_i / \partial r_i ) - \Sigma ( \partial p_i /
\partial r_i ) = 0
\end{eqnarray}
further, if the last term is equal to zero, we have
\begin{eqnarray}
\Sigma ( \partial P_i / \partial r_i ) = 0
\end{eqnarray}

The above states that the gravitational strength and adjoint
strength have an influence on continuity equation of adjoint mass,
although $\Sigma ( b_{gj} P_j -  B_{gj} p_j ) / v_0$ is usually very
tiny when field are weak. The continuity equation of adjoint mass is
the invariant under the octonion coordinate transformation.

Comparing Eq.(20) with Eq.(41), we find that the mass continuity
equation Eq.(21) and continuity equation of adjoint mass Eq.(42)
can't be effective at the same time. That means that some invariants
will not be effective simultaneously in the gravitational field and
adjoint field.

\subsubsection{Conservation of adjoint spin}

In the octonion space, a new physical quantity $\mathbb{L}_g =
\mathbb{L} \circ \emph{\textbf{I}}_0^*$ can be defined from Eq.(24).
\begin{eqnarray}
\mathbb{L}_g = L_0 + \Sigma (L_j \emph{\textbf{i}}_j ) - \Sigma (l_i
\emph{\textbf{I}}_i )
\end{eqnarray}
where, $L_0 = (r_0 + k_{rx} x_0 ) P_0 - \Sigma \left\{(r_j + k_{rx}
x_j ) P_j \right\} + \Sigma \left\{(R_i + k_{rx} X_i ) p_i \right\}
$ .

When the octonion coordinate system rotates, we have the angular
momentum density $\mathbb{L}'_g = \Sigma ( L'_i \emph{\textbf{i}}'_i
- l'_i \emph{\textbf{I}}'_i )$ from Eq.(6). Under the coordinate
transformation, the scalar part of $\mathbb{L}_g$ deduces the
conservation of adjoint spin.
\begin{eqnarray}
L_0 = L'_0
\end{eqnarray}

The above means that the adjoint spin density $L_0$ is an invariant
in the case for coexistence of the gravitational field and its
adjoint field, under the octonion coordinate transformation.

\subsubsection{Conservation of adjoint energy}

In the octonion space, a new physical quantity $\mathbb{W}_g =
\mathbb{W} \circ \emph{\textbf{I}}_0^*$ can be defined from Eq.(26).
\begin{eqnarray}
\mathbb{W}_g = W_0 + \Sigma (W_j \emph{\textbf{i}}_j ) - \Sigma (w_i
\emph{\textbf{I}}_i)
\end{eqnarray}
where, the scalar part $W_0 = v_0 \partial L_0 / \partial r_0 - v_0
\Sigma ( \partial L_j / \partial r_j ) + v_0 \Sigma ( \partial l_i /
\partial R_i ) - \Sigma ( b_{gj} L_j - B_{gj} l_j ) $.

When the octonion coordinate system rotates, we have the energy
density $\mathbb{W}'_g = \Sigma ( W'_i \emph{\textbf{i}}'_i - w'_i
\emph{\textbf{I}}'_i )$ from Eq.(6). Under the coordinate
transformation, the scalar part of $\mathbb{W}_g$ remains unchanged.
And then, we can obtain the conservation of adjoint energy.
\begin{eqnarray}
W_0 = W'_0
\end{eqnarray}

When the right side is equal to zero in the above, we have the
continuity equation of adjoint spin in the case for coexistence of
the adjoint field and gravitational field.
\begin{eqnarray}
&& \partial L_0 / \partial r_0 - \Sigma ( \partial L_j /
\partial r_j )
\nonumber\\
&& + \Sigma ( \partial l_i / \partial r_i ) - \Sigma ( b_{gj} L_j -
B_{gj} l_j ) / v_0 = 0
\end{eqnarray}

If the last term is neglected, the above is reduced to
\begin{eqnarray}
\partial L_0 / \partial r_0 - \Sigma ( \partial L_j / \partial r_j )
+ \Sigma ( \partial l_i / \partial r_i ) = 0
\end{eqnarray}
further, if the last term is equal to zero, we have
\begin{eqnarray}
\partial L_0 / \partial r_0 - \Sigma ( \partial L_j / \partial r_j ) = 0
\end{eqnarray}

The above means that the adjoint energy density $W_0$ is an
invariant in the case for coexistence of gravitational field and its
adjoint field, under the octonion coordinate transformation.

\begin{table}[t]
\caption{\label{tab:table1}The definitions and adjoint invariants of
physical quantities in the gravitational field with its adjoint
field in the octonion space.}
\begin{ruledtabular}
\begin{tabular}{llll}
$ definition $                           & $ invariant$       &    $conservation$                         \\
\hline
$\mathbb{R}\circ\emph{\textbf{I}}_0^*$   & $R_0 = R'_0$       &    $invariable~scalar$                    \\
$\mathbb{V}\circ\emph{\textbf{I}}_0^*$   & $V_0 = V'_0$       &    $invariable~speed~of~light-like$       \\
$\mathbb{X}\circ\emph{\textbf{I}}_0^*$   & $X_0 = X'_0$       &    $invariable~scalar$                    \\
$\mathbb{A}\circ\emph{\textbf{I}}_0^*$   & $A_0 = A'_0$       &    $invariable~adjoint~scalar~potential$  \\
$\mathbb{B}\circ\emph{\textbf{I}}_0^*$   & $B_0 = B'_0$       &    $invariable~adjoint~field~gauge$       \\
$\mathbb{P}\circ\emph{\textbf{I}}_0^*$   & $P_0 = P'_0$       &    $conservation~of~adjoint~mass$         \\
$\mathbb{F}\circ\emph{\textbf{I}}_0^*$   & $F_0 = F'_0$       &    $continuity~equation~of~adjoint~mass~$ \\
$\mathbb{L}\circ\emph{\textbf{I}}_0^*$   & $L_0 = L'_0$       &    $conservation~of~adjoint~spin$         \\
$\mathbb{W}\circ\emph{\textbf{I}}_0^*$   & $W_0 = W'_0$       &    $conservation~of~adjoint~energy$       \\
$\mathbb{N}\circ\emph{\textbf{I}}_0^*$   & $N_0 = N'_0$       &    $conservation~of~adjoint~power$        \\
\end{tabular}
\end{ruledtabular}
\end{table}

\subsubsection{Conservation of adjoint power}

In the octonion space, a new physical quantity $\mathbb{N}_g =
\mathbb{N} \circ \emph{\textbf{I}}_0^*$ can be defined from Eq.(31).
\begin{eqnarray}
\mathbb{N}_g = N_0 + \Sigma (N_j \emph{\textbf{i}}_j ) - \Sigma (n_i
\emph{\textbf{I}}_i)
\end{eqnarray}
where, the scalar $N_0 = v_0 \Sigma ( \partial W_i / \partial r_i )
- v_0 \Sigma ( \partial w_i / \partial R_i ) + \Sigma ( b_{gj} W_j -
B_{gj} w_j ) $.

When the octonion coordinate system rotates, we have the adjoint
power density $\mathbb{N}'_g = \Sigma ( N'_i \emph{\textbf{i}}'_i -
n'_i \emph{\textbf{I}}'_i )$ from Eq.(6). Under the coordinate
transformation, the scalar part of $\mathbb{N}_g$ remains unchanged
by the above. And then we obtain the conservation of adjoint power
as follows.
\begin{eqnarray}
N_0 = N'_0
\end{eqnarray}

When the right side is equal to zero in the above, we have the
continuity equation of adjoint energy in the case for coexistence of
the adjoint field and gravitational field.
\begin{eqnarray}
\Sigma \left\{ \partial ( W_i - w_i ) / \partial r_i \right\} +
\Sigma ( b_{gj} W_j - B_{gj} w_j ) / v_0 = 0
\end{eqnarray}

If the last term is neglected, the above is reduced to
\begin{eqnarray}
\Sigma ( \partial W_i / \partial r_i ) - \Sigma ( \partial w_i /
\partial r_i ) = 0
\end{eqnarray}
further, if the last term is equal to zero, we have
\begin{eqnarray}
\Sigma ( \partial W_i / \partial r_i ) = 0
\end{eqnarray}

The above means the adjoint power density $N_0$ is the invariant in
the case for coexistence of the gravitational field and adjoint
field, under the octonion coordinate transformation.

\section{Electromagnetic field}

Making use of the octonion operator, the electromagnetic field
demonstrated by the octonion operator will also generate an adjoint
field. The source of adjoint field includes the adjoint charge and
adjoint electric current. Similarly, the adjoint charge and its
movement can not be observed by usual experiments. However, when the
adjoint charge is combined with the ordinary charge to become the
charged particles, their movements will be accompanied by some
mechanical or electric effects. And this kind of adjoint charge may
be considered as one kind of candidate for dark matter
\cite{zwicky}.

The electromagnetic field and its adjoint field both can be
demonstrated by the quaternions also, although they are quite
different from each other indeed.

With the invariant property of octonions, we find that the adjoint
charge, adjoint mass, velocity curl, and field strength have the
influence on some conservation laws in the electromagnetic field,
under the octonion coordinate transformation.

In some cases, the adjoint charge is combined with the ordinary
charge to become one sort of particle. Further, the ordinary mass
$m$, adjoint mass $\bar{m}$, ordinary charge $q$, and adjoint charge
$\bar{q}$ can be combined together to become another sort of
particle. Therefore we can measure their various characteristics,
and have following relation.
\begin{eqnarray}
R_i \emph{\textbf{I}}_i = r_i \emph{\textbf{i}}_i \circ
\emph{\textbf{I}}_0~;~ V_i \emph{\textbf{I}}_i = v_i
\emph{\textbf{i}}_i \circ \emph{\textbf{I}}_0~.
\nonumber
\end{eqnarray}

\subsection{Invariants in electromagnetic field}

\subsubsection{Potential and strength}

The electromagnetic field potential is
\begin{eqnarray}
\mathbb{A}_e = \Sigma (A_i \emph{\textbf{I}}_i)~.
\end{eqnarray}

The electromagnetic potential are combined with the gravitational
potential to become the field potential $\mathbb{A} = \mathbb{A}_g +
k_a \mathbb{A}_e $, with $k_a$ being the coefficient.

The field strength $\mathbb{B} = \Sigma (b_i \emph{\textbf{i}}_i) +
\Sigma (B_i \emph{\textbf{I}}_i)$ consists of the gravitational
strength $\mathbb{B}_g$ and the electromagnetic strength
$\mathbb{B}_e$ , with $k_b$ being the coefficient.
\begin{eqnarray}
\mathbb{B} = \lozenge \circ \mathbb{A} = \mathbb{B}_g + k_b
\mathbb{B}_e
\end{eqnarray}

The strength $\mathbb{B}_e = \Sigma (b_{ei} \emph{\textbf{i}}_i) +
\Sigma (B_{ei} \emph{\textbf{I}}_i)$ consists of the electromagnetic
strength $\mathbb{B}_{eg}$ and adjoint strength $\mathbb{B}_{ee}$ .
\begin{eqnarray}
\mathbb{B}_e = \lozenge \circ \mathbb{A}_e = \mathbb{B}_{eg} +
\mathbb{B}_{ee}
\end{eqnarray}
where, $\mathbb{B}_{ee} = \Sigma (b_{ei} \emph{\textbf{i}}_i)$,
$\mathbb{B}_{eg} = \Sigma (B_{ei} \emph{\textbf{I}}_i)$; $b_{ei}$
and $B_{ei}$ are all real.

In the above equation, we choose the following gauge conditions to
simplify succeeding calculation.
\begin{eqnarray}
&& \partial A_0 / \partial r_0 - \Sigma (\partial A_j /
\partial r_j) = 0 \nonumber\\
&& \partial A_0 / \partial R_0 + \Sigma (\partial A_j /
\partial R_j) = 0 \nonumber
\end{eqnarray}

The adjoint field strength $\mathbb{B}_{ee}$ in Eq.(59) includes two
parts, $\textbf{g}_e = ( g_{e01} , g_{e02} , g_{e03} ) $ and
$\textbf{b}_e = ( g_{e23} , g_{e31} , g_{e12} )$,
\begin{eqnarray}
\textbf{g}_e/v_0 = && \emph{\textbf{i}}_1 ( \partial_0 A_1 -
\partial_1 A_0 ) + \emph{\textbf{i}}_2 ( \partial_0 A_2 - \partial_2
A_0 )
\nonumber\\
&& + \emph{\textbf{i}}_3 ( \partial_0 A_3 - \partial_3 A_0 )
\\
\textbf{b}_e = && \emph{\textbf{i}}_1 ( \partial_3 A_2 -
\partial_2 A_3 ) + \emph{\textbf{i}}_2 ( \partial_1 A_3 - \partial_3
A_1 )
\nonumber\\
&& + \emph{\textbf{i}}_3 ( \partial_2 A_1 - \partial_1 A_2 )
\end{eqnarray}
simultaneously, the electromagnetic field strength $\mathbb{B}_{eg}$
involves two components, $\textbf{E}_e = ( B_{e01} , B_{e02} ,
B_{e03} ) $ and $\textbf{B}_e = ( B_{e23} , B_{e31} , B_{e12} )$ .
\begin{eqnarray}
\textbf{E}_e/v_0 = && \emph{\textbf{I}}_1 ( \partial_0 A_1 +
\partial_1 A_0 ) + \emph{\textbf{I}}_2 ( \partial_0 A_2 + \partial_2
A_0 )
\nonumber\\
&& + \emph{\textbf{I}}_3 ( \partial_0 A_3 + \partial_3 A_0 )
\\
\textbf{B}_e = && \emph{\textbf{I}}_1 ( \partial_3 A_2 -
\partial_2 A_3 ) + \emph{\textbf{I}}_2 ( \partial_1 A_3 - \partial_3
A_1 )
\nonumber\\
&& + \emph{\textbf{I}}_3 ( \partial_2 A_1 - \partial_1 A_2 )
\end{eqnarray}

The electric current density $\mathbb{S}_{eg} = q \mathbb{V}_g \circ
\emph{\textbf{I}}_0$ is the source of electromagnetic field, and its
adjoint electric current density $\mathbb{S}_{ee} = \bar{q}
\mathbb{V}_g $ is that of adjoint field. They combine together to
become the field source $\mathbb{S}_e$ .

In the octonion space, the electromagnetic source $\mathbb{S}_e$ can
be combined with gravitational source $\mathbb{S}_g$ to become the
source $\mathbb{S}$ .
\begin{eqnarray}
\mu \mathbb{S} = && - ( \mathbb{B}/v_0 + \lozenge)^* \circ
\mathbb{B}
\nonumber\\
= && \mu_{gg} \mathbb{S}_{gg} + \mu_{ge} \mathbb{S}_{ge}  -
\mathbb{B}^* \circ \mathbb{B}/v_0
\nonumber\\
&& + k_b (\mu_{ee} \mathbb{S}_{ee} + \mu_{eg} \mathbb{S}_{eg})
\end{eqnarray}
where, $k_b^2 = \mu_{gg} /\mu_{eg}$; $\mu_{gg}$, $\mu_{ge}$,
$\mu_{ee}$, and $\mu_{eg}$ are the coefficients.

The $\mathbb{B}^* \circ \mathbb{B}/(2\mu_{gg})$ is the field energy
density.
\begin{eqnarray}
\mathbb{B}^* \circ \mathbb{B}/ \mu_{gg} = \mathbb{B}_g^* \circ
\mathbb{B}_g / \mu_{gg} + \mathbb{B}_e^* \circ \mathbb{B}_e /
\mu_{eg}
\end{eqnarray}

The above means that the electromagnetic field and its adjoint field
make a contribution to the gravitational mass also in the octonion
space.

\subsubsection{Conservation of mass}

In the electromagnetic field, gravitational field and their adjoint
fields, the linear momentum density $\mathbb{P} = \mu \mathbb{S} /
\mu_{gg}$ is written as
\begin{eqnarray}
\mathbb{P} = && \widehat{m} v_0 + \Sigma (m v_j \emph{\textbf{i}}_j
) + \Sigma ( M_g V_i \emph{\textbf{i}}_i \circ \emph{\textbf{I}}_0 )
\nonumber\\
&& + \Sigma ( M_q V_i \emph{\textbf{i}}_i \circ \emph{\textbf{I}}_0
) + \Sigma ( M_e v_i \emph{\textbf{i}}_i )
\end{eqnarray}
where, $\widehat{m} = m - (\mathbb{B}^* \circ \mathbb{B} / \mu_{gg}
)/v_0^2 $ ; $M_q = q k_b \mu_{eg}/\mu_{gg}$ ; $M_e = \bar{q} k_b
\mu_{ee}/\mu_{gg}$ .

The above means that the gravitational mass density $\widehat{m}$ is
changed with all four kinds of field strengthes in the
electromagnetic field, gravitational field, and their two kinds of
adjoint fields.

From Eq.(6), we have one linear momentum density, $\mathbb{P}'
(p'_0, p'_1, p'_2, p'_3, P'_0, P'_1, P'_2, P'_3)$, when the octonion
coordinate system is rotated. And we obtain the invariant equation
from Eqs.(7) and (66).
\begin{eqnarray}
(\widehat{m} + M_e ) v_0 = (\widehat{m}' + M'_e) v'_0
\end{eqnarray}

Under Eqs.(3), (7), and (67), we find the gravitational mass density
$(\widehat{m} + M_e)$ remains unchanged.
\begin{eqnarray}
\widehat{m} + M_e = \widehat{m}' + M'_e
\end{eqnarray}

The above means that if we choose the definitions of velocity and
linear momentum, the inertial mass density $(m + M_e)$ and
gravitational mass density $(\widehat{m} + M_e)$ will keep unchanged
respectively, under the octonion coordinate transformation in Eq.(6)
in the electromagnetic field, gravitational field, and their adjoint
fields.

\subsubsection{Mass continuity equation}

In the electromagnetic field, gravitational field, and their adjoint
fields, the applied force density $\mathbb{F} = \Sigma (f_i
\emph{\textbf{i}}_i ) + \Sigma (F_i \emph{\textbf{I}}_i )$ is
defined from the linear momentum density $\mathbb{P} = \Sigma (p_i
\emph{\textbf{i}}_i ) + \Sigma (P_i \emph{\textbf{I}}_i )$ in
Eq.(66).
\begin{eqnarray}
\mathbb{F} = v_0 (\mathbb{B}/v_0 + \lozenge )^* \circ \mathbb{P}
\end{eqnarray}
where, the scalar $f_0 = v_0 \Sigma (\partial p_i / \partial r_i) +
v_0 \Sigma ( \partial P_i / \partial R_i ) + \Sigma ( b_{gj} p_j +
B_{gj} P_j + k_b b_{ej} p_j + k_b B_{ej} P_j ) $.

When the coordinate system rotates, we have the new force density
$\mathbb{F}' (f'_0, f'_1, f'_2, f'_3, F'_0, F'_1, F'_2, F'_3)$.

By Eq.(7), we have
\begin{eqnarray}
f_0 = f'_0
\end{eqnarray}

When $f'_0 = 0$ in the above, we have the conservation of mass in
the case for coexistence of the electromagnetic field, gravitational
field, and their adjoint fields.
\begin{eqnarray}
&& \Sigma \left\{ \partial (p_i + P_i )/ \partial r_i \right\}  +
\Sigma ( b_{gj} p_j + B_{gj} P_j ) / v_0
\nonumber\\
&& + \Sigma k_b ( b_{ej} p_j + B_{ej} P_j ) / v_0 = 0
\end{eqnarray}

If the $b_{gi} = B_{gi} = b_{ei} = B_{ei} = 0$, the above will be
reduced to the following equation.
\begin{eqnarray}
\Sigma ( \partial p_j / \partial r_j ) + \Sigma ( \partial P_j /
\partial r_j ) = 0
\end{eqnarray}
further, if the last term can be neglected, we have
\begin{eqnarray}
\Sigma (\partial p_i / \partial r_i) = 0
\end{eqnarray}

In case of we choose the definitions of applied force and velocity
in the electromagnetic field, gravitational field, and their adjoint
fields, the conservation of mass will be the invariant under the
octonion transformation in Eq.(6). The above states also that four
kinds of field strengthes, adjoint mass, and adjoint charge have a
tiny influence on conservation of mass, although the impact is
usually very small when the fields are weak.

\subsubsection{Conservation of spin}

The angular momentum density $\mathbb{L} = \Sigma (l_i
\emph{\textbf{i}}_i ) + \Sigma (L_i \emph{\textbf{I}}_i )$ is
defined from the radius vector $\mathbb{R}$, physical quantity
$\mathbb{X}$, and linear momentum density $\mathbb{P}$ in the
octonion space.
\begin{eqnarray}
\mathbb{L} = (\mathbb{R} + k_{rx} \mathbb{X}) \circ \mathbb{P}
\end{eqnarray}
where, $l_0$ is considered as the spin angular momentum density in
the gravitational field and adjoint field; $l_0 = (r_0 + k_{rx} x_0
) p_0 - \Sigma \left\{ (r_j + k_{rx} x_j ) p_j \right\} - \Sigma
\left\{ (R_i + k_{rx} X_i ) P_i \right\}$; $k_{rx}$ is the
coefficient.

When the octonion coordinate system rotates, we have the new angular
momentum density $\mathbb{L}' = \Sigma ( l'_i \emph{\textbf{i}}'_i +
L'_i \emph{\textbf{I}}'_i )$. Under the octonion coordinate
transformation, the spin density remains unchanged from Eq.(7).
\begin{eqnarray}
l_0 = l'_0
\end{eqnarray}

The above means that the space, time, adjoint fields,
electromagnetic field, and gravitational field have small influence
on orbital angular momentum and spin angular momentum. The spin
angular momentum density $l_0$ will change with time, although $l_0$
is an invariant under the octonion transformation.

\begin{table}[b]
\caption{\label{tab:table1}The definitions and the mechanics
invariants of electromagnetic field and gravitational field with
their adjoint fields in the octonion space.}
\begin{ruledtabular}
\begin{tabular}{lll}
$definition$    &    $invariant $    &    $ meaning $                            \\
\hline
$\mathbb{R}$    &    $r_0 = r'_0 $   &    $Galilean~invariant$                   \\
$\mathbb{V}$    &    $v_0 = v'_0$    &    $invariable~speed~of~light$            \\
$\mathbb{A}$    &    $a_0 = a'_0$    &    $invariable~scalar~potential$          \\
$\mathbb{B}$    &    $b_0 = b'_0$    &    $invariable~gauge$                     \\
$\mathbb{P}$    &    $p_0 = p'_0$    &    $invariable~mass~density$              \\
$\mathbb{F}$    &    $f_0 = f'_0$    &    $conservation~of~mass$                 \\
$\mathbb{L}$    &    $l_0 = l'_0$    &    $invariable~spin~density$              \\
$\mathbb{W}$    &    $w_0 = w'_0$    &    $invariable~energy~density$            \\
$\mathbb{N}$    &    $n_0 = n'_0$    &    $conservation~of~energy$               \\
\end{tabular}
\end{ruledtabular}
\end{table}

\subsubsection{Conservation of energy}

The total energy density $\mathbb{W} = \Sigma (w_i
\emph{\textbf{i}}_i ) + \Sigma (W_i \emph{\textbf{I}}_i )$ is
defined from the angular momentum density $\mathbb{L}$.
\begin{eqnarray}
\mathbb{W} = v_0 ( \mathbb{B}/v_0 + \lozenge) \circ \mathbb{L}
\end{eqnarray}
where, the scalar part $w_0 = v_0 \partial l_0 / \partial r_0 - v_0
\Sigma (\partial l_j / \partial r_j) - v_0 \Sigma ( \partial L_i /
\partial R_i ) - \Sigma ( b_{gj} l_j + B_{gj} L_j + k_b b_{ej} l_j +
k_b B_{ej} L_j ) $.

When the coordinate system rotates, we have the new energy density
$\mathbb{W}' = \Sigma ( w'_i \emph{\textbf{i}}'_i + W'_i
\emph{\textbf{I}}'_i )$. Under the octonion transformation, the
scalar part of total energy density is the energy density and
remains unchanged by Eq.(7).
\begin{eqnarray}
w_0 = w'_0
\end{eqnarray}

In some special cases, the right side is equal to zero. We obtain
the conservation of spin angular momentum.
\begin{eqnarray}
&& - \Sigma ( b_{gj} l_j + B_{gj} L_j + k_b b_{ej} l_j + k_b B_{ej}
L_j ) / v_0
\nonumber\\
&& + \partial l_0 / \partial r_0 - \Sigma (\partial l_j /
\partial r_j) - \Sigma ( \partial L_i /
\partial r_i ) = 0
\end{eqnarray}

If the first term is zero, the above is reduced to
\begin{eqnarray}
\partial l_0 / \partial r_0 - \Sigma (\partial l_j /
\partial r_j) - \Sigma ( \partial L_i /
\partial r_i ) = 0
\end{eqnarray}
further, if last term is zero, we have
\begin{eqnarray}
\partial l_0 / \partial r_0 - \Sigma (\partial l_j /
\partial r_j) = 0
\end{eqnarray}

The above means the energy density $w_0$ is variable with time in
the case for coexistence of the electromagnetic field, gravitational
field, and their adjoint fields, because the adjoint mass, adjoint
charge, velocity, and strength have the influence on the angular
momentum density. While the scalar $w_0$ is the invariant under the
octonion transformation from Eq.(7).

\subsubsection{Conservation of power}

In the electromagnetic field and gravitational field with their
adjoint fields, the external power density $\mathbb{N}= \Sigma (n_i
\emph{\textbf{i}}_i ) + \Sigma (N_i \emph{\textbf{I}}_i )$ can be
defined from the total energy density $\mathbb{W}$ .
\begin{eqnarray}
\mathbb{N} = v_0 ( \mathbb{B}/v_0 + \lozenge)^* \circ \mathbb{W}
\end{eqnarray}
where, the scalar $n_0 = v_0 \Sigma (\partial w_i / \partial r_i) +
v_0 \Sigma ( \partial W_i / \partial R_i ) + \Sigma ( b_{gj} w_j +
B_{gj} W_j + k_b b_{ej} w_j + k_b B_{ej} W_j ) $.

When the coordinate system rotates, we have the new external power
density $\mathbb{N}' = \Sigma ( n'_i \emph{\textbf{i}}'_i + N'_i
\emph{\textbf{I}}'_i )$. Under the octonion coordinate
transformation, the scalar part of external power density is the
power density and remains unchanged by Eq.(7).
\begin{eqnarray}
n_0 = n'_0
\end{eqnarray}

In a special case, the right side is equal to zero. And then, we
obtain the conservation of energy.
\begin{eqnarray}
&& \Sigma \left\{ \partial (w_i + W_i )/
\partial r_i \right\} + \Sigma ( b_{gj} w_j + B_{gj} W_j ) / v_0
\nonumber\\
&& + \Sigma k_b ( b_{ej} w_j + B_{ej} W_j ) / v_0 = 0
\end{eqnarray}

If last two term are zeros, the above is reduced to
\begin{eqnarray}
\Sigma ( \partial w_i / \partial r_i ) + \Sigma ( \partial W_i /
\partial r_i ) = 0
\end{eqnarray}
further, if last term is equal to zero, we have
\begin{eqnarray}
\Sigma ( \partial w_i / \partial r_i ) = 0
\end{eqnarray}

The above means that the power density $n_0$ will be variable in the
case for coexistence of the electromagnetic field, gravitational
field, and their adjoint fields, although the $n_0$ is the invariant
under the octonion transformation. And the adjoint mass, adjoint
charge, field strength, and torque density etc. have a few influence
on the energy continuity equation in the octonion space.

\subsection{Invariants in electromagnetic adjoint field}

\subsubsection{Conservation of charge}

In the adjoint field, a new physical quantity $\mathbb{P}_q =
\mathbb{P} \circ \emph{\textbf{I}}_0^* $ can be defined from
Eq.(66).
\begin{eqnarray}
\mathbb{P}_q = P_0 + \Sigma (P_j \emph{\textbf{i}}_j ) - \left\{ p_0
\emph{\textbf{I}}_0 + \Sigma (p_j \emph{\textbf{I}}_j ) \right\}
\end{eqnarray}
where, $P_0 = (M_g + M_q ) V_i$ .

By Eq.(6), we have the linear momentum density, $\mathbb{P}'_q =
\Sigma ( P'_i \emph{\textbf{i}}'_i - p'_i \emph{\textbf{I}}'_i )$,
when the octonion coordinate system is rotated. Under the octonion
coordinate transformation, the scalar part of $\mathbb{P}_q$ remains
unchanged.
\begin{eqnarray}
(M_g + M_q ) V_0 = (M'_g + M'_q ) V'_0
\end{eqnarray}

By means of Eqs.(3), (7), and the above, we obtain the conservation
of charge as follows. And $(M_g + M_q )$ is a scalar invariant,
which is a function of the adjoint mass density $\bar{m}$ and
ordinary charge $q$ .
\begin{eqnarray}
M_g + M_q = M'_g + M'_q
\end{eqnarray}

The above means if we emphasize definitions of velocity and linear
momentum, the charge density $(M_g + M_q )$ will remain the same,
under the coordinate transformation in the electromagnetic field,
gravitational field, and their adjoint fields.

\subsubsection{Charge continuity equation}

In the octonion space, a new physical quantity $\mathbb{F}_q =
\mathbb{F} \circ \emph{\textbf{I}}_0^*$ can be defined from Eq.(69).
\begin{eqnarray}
\mathbb{F}_q = F_0 + \Sigma (F_j \emph{\textbf{i}}_j ) - \Sigma (f_i
\emph{\textbf{I}}_i )
\end{eqnarray}
where, the scalar $F_0 = v_0 \Sigma ( \partial P_i / \partial r_i )
- v_0 \Sigma ( \partial p_i / \partial R_i ) + \Sigma ( b_{gj} P_j -
B_{gj} p_j + k_b b_{ej} P_j - k_b B_{ej} p_j ) $.

When the octonion coordinate system rotates, we have the applied
force density $\mathbb{F}'_q = \Sigma ( F'_i \emph{\textbf{i}}'_i -
f'_i \emph{\textbf{I}}'_i )$. Under the coordinate transformation,
the scalar part of $\mathbb{F}_q$ remains unchanged.
\begin{eqnarray}
F_0 = F'_0
\end{eqnarray}

When the right side is equal to zero, we have the charge continuity
equation in the case for coexistence of the gravitational field,
electromagnetic field, and adjoint fields.
\begin{eqnarray}
&& \Sigma \left\{ \partial ( P_i - p_i ) / \partial r_i \right\} +
\Sigma ( b_{gj} P_j - B_{gj} p_j ) / v_0
\nonumber\\
&& + \Sigma k_b ( b_{ej} P_j - B_{ej} p_j ) = 0
\end{eqnarray}

If last two terms are zeros, the above is reduced to
\begin{eqnarray}
\Sigma ( \partial P_i / \partial r_i ) - \Sigma ( \partial p_i /
\partial r_i ) = 0
\end{eqnarray}
further, if the last term is equal to zero, we have
\begin{eqnarray}
\Sigma ( \partial P_i / \partial r_i ) = 0
\end{eqnarray}

The above states that the electromagnetic strength, gravitational
strength, and adjoint field strengthes have an effect on the charge
continuity equation, although the impact is usually very tiny when
fields are weak. And the charge continuity equation is an invariant
under the octonion coordinate transformation.

\subsubsection{Conservation of spin magnetic moment}

In the octonion space, a new physical quantity $\mathbb{L}_q =
\mathbb{L} \circ \emph{\textbf{I}}_0^*$ can be defined from Eq.(74).
\begin{eqnarray}
\mathbb{L}_q = L_0 + \Sigma (L_j \emph{\textbf{i}}_j ) - \Sigma (l_i
\emph{\textbf{I}}_i )
\end{eqnarray}
where, $L_0 = (r_0 + k_{rx} x_0 ) P_0 - \Sigma \left\{(r_j + k_{rx}
x_j ) P_j \right\} + \Sigma \left\{(R_i + k_{rx} X_i ) p_i \right\}
$ .

When the octonion coordinate system rotates, we have the angular
momentum density $\mathbb{L}'_q = \Sigma ( L'_i \emph{\textbf{i}}'_i
- l'_i \emph{\textbf{I}}'_i )$ from Eq.(6). Under the octonion
coordinate transformation, the scalar part of $\mathbb{L}_q$ deduces
the conservation of spin magnetic moment.
\begin{eqnarray}
L_0 = L'_0
\end{eqnarray}

The above states the spin magnetic moment density $L_0$ is an
invariant in the case for coexistence of the electromagnetic field,
gravitational field and their adjoint fields, under the octonion
coordinate transformation.

\subsubsection{Conservation of energy-like}

In the octonion space, a new physical quantity $\mathbb{W}_q =
\mathbb{W} \circ \emph{\textbf{I}}_0^*$ can be defined from Eq.(76).
\begin{eqnarray}
\mathbb{W}_q = W_0 + \Sigma (W_j \emph{\textbf{i}}_j ) - \Sigma (w_i
\emph{\textbf{I}}_i)
\end{eqnarray}
where, the scalar part $W_0 = v_0 \partial L_0 / \partial r_0 - v_0
\Sigma ( \partial L_j / \partial r_j ) + v_0 \Sigma (
\partial l_i / \partial R_i ) - \Sigma ( b_{gj} L_j - B_{gj} l_j +
k_b b_{ej} L_j - k_b B_{ej} l_j ) $.

When the octonion coordinate system rotates, we have the energy
density $\mathbb{W}'_q = \Sigma ( W'_i \emph{\textbf{i}}'_i - w'_i
\emph{\textbf{I}}'_i )$ from Eq.(6). Under the coordinate
transformation, the scalar part of $\mathbb{W}_q$ remains unchanged.
Therefore, we can obtain the conservation of energy-like.
\begin{eqnarray}
W_0 = W'_0
\end{eqnarray}

When the right side is equal to zero in the above, we have the
continuity equation of spin magnetic moment in the case for
coexistence of the electromagnetic field, gravitational field and
their adjoint fields.
\begin{eqnarray}
&& - \Sigma ( b_{gj} L_j - B_{gj} l_j + k_b b_{ej} L_j - k_b B_{ej}
l_j ) / v_0
\nonumber\\
&& + \partial L_0 / \partial r_0 - \Sigma ( \partial L_j /
\partial r_j ) + \Sigma ( \partial l_i / \partial r_i ) = 0
\end{eqnarray}

If the first term is zero, the above is reduced to
\begin{eqnarray}
\partial L_0 / \partial r_0 - \Sigma ( \partial L_j /
\partial r_j ) + \Sigma ( \partial l_i / \partial r_i ) = 0
\end{eqnarray}
further, if the last term is equal to zero, we have
\begin{eqnarray}
\partial L_0 / \partial r_0 - \Sigma ( \partial L_j /
\partial r_j ) = 0
\end{eqnarray}

The above means that the energy-like density $W_0$ is an invariant
in the case for coexistence of the electromagnetic field,
gravitational field and their adjoint fields, under the octonion
coordinate transformation.

\begin{table}[t]
\caption{\label{tab:table1}The definitions and their adjoint
invariants of the physical quantities in the electromagnetic field,
gravitational field with their adjoint fields.}
\begin{ruledtabular}
\begin{tabular}{llll}
$ definition $                           & $ invariant$       &    $conservation$                      \\
\hline
$\mathbb{R}\circ\emph{\textbf{I}}_0^*$   & $R_0 = R'_0$       &    $invariable~scalar$                 \\
$\mathbb{V}\circ\emph{\textbf{I}}_0^*$   & $V_0 = V'_0$       &    $invariable~speed~of~light-like$    \\
$\mathbb{X}\circ\emph{\textbf{I}}_0^*$   & $X_0 = X'_0$       &    $invariable~scalar$                 \\
$\mathbb{A}\circ\emph{\textbf{I}}_0^*$   & $A_0 = A'_0$       &    $electric~scalar~potential$         \\
$\mathbb{B}\circ\emph{\textbf{I}}_0^*$   & $B_0 = B'_0$       &    $electromagnetic~gauge$             \\
$\mathbb{P}\circ\emph{\textbf{I}}_0^*$   & $P_0 = P'_0$       &    $conservation~of~charge$            \\
$\mathbb{F}\circ\emph{\textbf{I}}_0^*$   & $F_0 = F'_0$       &    $charge~continuity~equation$        \\
$\mathbb{L}\circ\emph{\textbf{I}}_0^*$   & $L_0 = L'_0$       &    $spin~magnetic~moment$              \\
$\mathbb{W}\circ\emph{\textbf{I}}_0^*$   & $W_0 = W'_0$       &    $conservation~of~energy-like$       \\
$\mathbb{N}\circ\emph{\textbf{I}}_0^*$   & $N_0 = N'_0$       &    $conservation~of~power-like$        \\
\end{tabular}
\end{ruledtabular}
\end{table}

\subsubsection{Conservation of power-like}

In the octonion space, a new physical quantity $\mathbb{N}_q =
\mathbb{N} \circ \emph{\textbf{I}}_0^*$ can be defined from Eq.(81).
\begin{eqnarray}
\mathbb{N}_q = N_0 + \Sigma (N_j \emph{\textbf{i}}_j ) - \Sigma (n_i
\emph{\textbf{I}}_i)
\end{eqnarray}
where, the scalar $N_0 = v_0 \Sigma ( \partial W_i / \partial r_i )
- v_0 \Sigma ( \partial w_i / \partial R_i ) + \Sigma ( b_{gj} W_j -
B_{gj} w_j + k_b b_{ej} W_j - k_b B_{ej} w_j ) $.

When the octonion coordinate system rotates, we have the power-like
density $\mathbb{N}'_q = \Sigma ( N'_i \emph{\textbf{i}}'_i - n'_i
\emph{\textbf{I}}'_i )$ from Eq.(6). Under the coordinate
transformation, the scalar part of $\mathbb{N}_q$ remains unchanged
by the above. And then we obtain the conservation of power-like as
follows.
\begin{eqnarray}
N_0 = N'_0
\end{eqnarray}

When the right side is equal to zero in the above, we have the
continuity equation of energy-like in the case for coexistence of
the electromagnetic field, gravitational field and their adjoint
fields.
\begin{eqnarray}
&& \Sigma \left\{ \partial ( W_i - w_i ) / \partial r_i \right\} +
\Sigma ( b_{gj} W_j - B_{gj} w_j ) / v_0
\nonumber\\
&& + \Sigma k_b ( b_{ej} W_j - B_{ej} w_j ) / v_0 = 0
\end{eqnarray}

If last two terms are zeros, the above is reduced to
\begin{eqnarray}
\Sigma ( \partial W_i / \partial r_i ) - \Sigma ( \partial w_i /
\partial r_i ) = 0
\end{eqnarray}
further, if the last term is equal to zero, we have
\begin{eqnarray}
\Sigma ( \partial W_i / \partial r_i ) = 0
\end{eqnarray}

The above means that the power-like density $N_0$ is one scalar
invariant in the case for coexistence of the electromagnetic field,
gravitational field and their adjoint fields, under the octonion
coordinate transformation.

\section{CONCLUSIONS}

In the octonion space, the gravitational field described by the
octonion operator will generate one adjoint field. Similarly, the
electromagnetic field will be accompanied by its adjoint field.
These two sorts of adjoint fields will impact the scalar invariants
and conservation laws in the electromagnetic field and gravitational
field.

In the gravitational field with its gravitational adjoint field, the
gravitational mass density is changed with the gravitational
strength and adjoint field strength. And the mass continuity
equation will be changed with the field strength, velocity, and
adjoint mass. From the definitions of the angular momentum and
velocity, the spin density, energy density, and power density will
be variable for the influence of the adjoint field potential and
adjoint field strength. While the spin continuity equation and
energy continuity equation will be changed with the impact of the
velocity and adjoint field strength etc.

The gravitational mass density will be variable in the case for
coexistence of the gravitational field and electromagnetic field
with their adjoint fields. The gravitational mass density is changed
with the gravitational strength, electromagnetic strength,
gravitational adjoint field, and electromagnetic adjoint field. And
the mass continuity equation will be changed with the
electromagnetic field strength, velocity, and adjoint charge. The
spin density, energy density, and power density all will be variable
for the influence of the electromagnetic adjoint field, adjoint
field strength, and adjoint charge. Meanwhile the spin continuity
equation and energy continuity equation will be changed with the
impact of the adjoint charge and electromagnetic adjoint field
strength etc.

In the gravitational field and electromagnetic field with their
adjoint fields, there exist some electric invariants, which are
associated with ordinary charge and adjoint mass. These electric
invariants will be variable for the influence of the field potential
and field strength. And some conservation laws and scalar invariants
can not be effective simultaneously.

It should be noted that the study for some scalar invariants of
electromagnetic and gravitational adjoint fields examined only one
simple case with very weak field strength and low velocity in the
gravitational field and electromagnetic field with their adjoint
fields. Despite its preliminary character, this study can clearly
indicate the field strength and adjoint fields of the gravitational
field and electromagnetic field have the limited influence on the
scalar invariants. For the future studies, the related investigation
will concentrate on only the predictions of scalar invariants in the
strong adjoint field strength with high velocity in gravitational
field and electromagnetic field with their adjoint fields.

\begin{acknowledgments}
This project was supported partially by the National Natural Science
Foundation of China under grant number 60677039, Science \&
Technology Department of Fujian Province of China under grant number
2005HZ1020 and 2006H0092, and Xiamen Science \& Technology Bureau
of China under grant number 3502Z20055011.
\end{acknowledgments}

\end{document}